\documentclass[12pt,russian]{article}
 \usepackage{helvet}
 \usepackage[T1]{fontenc}
 \usepackage[cp1251]{inputenc}
 \usepackage{geometry}
 \usepackage{graphicx}
 \usepackage{setspace}
 \usepackage{babel}

 \def\gsim{\lower.4ex\hbox{$\;\buildrel >\over{\scriptstyle\sim}\;$}}
 \def\lsim{\lower.4ex\hbox{$\;\buildrel <\over{\scriptstyle\sim}\;$}}
 
 \def\bl{\par\vskip 12pt\noindent}
 \def\bll{\par\vskip 24pt\noindent}
 \def\blll{\par\vskip 36pt\noindent}

 \def\beg{\begin{eqnarray}}
 \def\ende{\end{eqnarray}}

 \def\aa{A\&A}
 \def\apj{ApJ}
 
 \def\an{Astron. Nachr.}
 
 \def\mnras{MNRAS}
 
 \def\sp{Solar Phys.}

\begin{document}

\vskip 1.0 cm

\begin{center}
{\bf Does the Babcock–Leighton Mechanism Operate on the Sun?}

\end{center}

\bl

\centerline{{\bf L.~L.~Kitchatinov}$^{1,2}$\footnote{E-mail:
kit@iszf.irk.ru}, {\bf S.~V.~Olemskoy}$^1$}

\bl

\begin{center}
$^1${\it Institute for Solar–Terrestrial Physics,
P.O. Box 291, Irkutsk, 664033, Russian Federation} \\
$^2${\it Pulkovo Astronomical Observatory, St. Petersburg, 196140,
Russian Federation}
\end{center}

\bll \hspace{0.8 truecm}
\parbox{15.0 truecm}{
{\bf Abstract}—The contribution of the Babcock–Leighton mechanism to
the generation of the Sun’s poloidal magnetic field is estimated
from sunspot data for three solar cycles. Comparison of the derived
quantities with the A-index of the large-scale magnetic field
suggests a positive answer to the question posed in the title of
this paper.

\bl {\bf DOI:} 10.1134/S0320010811080031

\bl Keywords: {\sl Sun: magnetic fields—sunspots—dynamo.}}

\blll

\reversemarginpar

\setlength{\baselineskip}{0.6 truecm}

 \centerline{INTRODUCTION}
 \bl

The dynamo theory explains the magnetic activity of the Sun by the
action of two main effects: the generation of a toroidal field from
a poloidal one by differential rotation (the $\Omega$-effect) and
the inverse transformation of the toroidal field to the poloidal one
by small-scale cyclonic motions (the $\alpha$-effect). The magnetic
field generation mechanism based on these two effects is called
$\alpha\Omega$-dynamo (see, e.g., Vainshtein et al. 1980).

The relatively simple $\Omega$-effect has been studied extensively
and its presence on the Sun is confirmed by observations. The
rotation of sunspots (Newton and Nunn 1951) and direct Doppler
measurements (Howard and Harvey 1970) determine the latitude
dependence of the rotation rate on the solar surface. Using this
dependence, helioseismology finds the angular velocity distribution
in the solar interior (Wilson et al. 1997; Schou et al. 1998). The
differential rotation changes little with time (LaBonte and Howard
1982), contains no small-scale inhomogeneities, and, in this sense,
is regular. Therefore, one may expect the toroidal field strength
and the related sunspot activity at solar maximum to be determined
by the poloidal field strength at the preceding minimum. This
circumstance was first pointed out by Schatten et al. (1978).
Makarov and Tlatov (2000) and Makarov et al. (2001) found a high
correlation between the $A$-index of the large-scale field at
activity minimum proposed by them and the amplitude of the
succeeding maximum. For three solar cycles for which direct
measurements of the polar field at activity minima are available,
the measured field strengths also correlate with the amplitudes of
the succeeding maxima (Jiang et al. 2007).

The situation with the $\alpha$-effect is completely different.
Several causes for the emergence of this effect are known. It can
appear due to density inhomogeneity in a turbulent rotating medium
(Parker 1955) and turbulence intensity inhomogeneity (Steenbeck et
al. 1966). Magnetostrophic waves at the base of the solar convection
zone can give rise to it (Schmitt 1987). Finally, the poloidal
magnetic field generation mechanism proposed by Babcock (1961) and
Leighton (1969) is also a variety of the $\alpha$-effect. The
Babcock–Leighton mechanism deserves special attention for two
reasons. First, in contrast to other varieties of the
$\alpha$-effect, it is not subject to the so-called catastrophic
quenching due to the conservation of magnetic helicity (Kitchatinov
and Olemskoy 2011a, 2011b). Therefore, it is quite possible that the
Babcock–Leighton mechanism is dominant in the generation of the
Sun’s poloidal field. Second, this mechanism is related to the
observed characteristics of solar active regions; therefore, its
contribution to the magnetic field generation can be estimated from
observational data. In this paper, we make such an estimate for
three solar cycles.

Sunspots are most likely associated with the emergence of toroidal
fields from deep layers of the Sun (see, e.g., Parker 1979). The
$\alpha$-effect transforms the toroidal fields into poloidal ones.
Therefore, one may expect the poloidal field at solar minimum to
correlate with the sunspot activity of the preceding cycle. However,
observations reveal no such correlation (Makarov and Tlatov 2000;
Makarov et al. 2001). This can be due to random fluctuations of the
$\alpha$-effect. Not only the toroidal field strength but also the
characteristics of the $\alpha$-effect responsible for the
transformation of this field to the poloidal one are important for
the poloidal field generation. Random fluctuations of the
$\alpha$-effect break the functional relation between the toroidal
and poloidal magnetic field components (Jiang et al. 2007; Moss et
al. 2008). Nevertheless, the contribution of the Babcock–Leighton
mechanism to the poloidal field generation, including the existing
fluctuations, can be estimated from sunspot data. In this paper, we
make such an estimate.

 \bll
 \centerline{THE METHOD}
 \bl

The Babcock–Leighton mechanism is associated with the so-called
Joy’s law for active regions on the Sun. The law asserts that the
preceding (leading in the rotational motion) sunspots of bipolar
groups are, on average, closer to the equator than the following
ones. Thus, on average, there is a (positive) tilt of the line
connecting the centers of opposite polarities to the solar parallels
(Fig. 1). The average tilt angle $\alpha$ increases with latitude,
suggesting that Joy’s law is related to the effect of the Coriolis
force on the rising magnetic loops (Wang and Sheeley 1989).

\begin{figure}[htb]
 \centerline{
 \includegraphics[width=9cm]{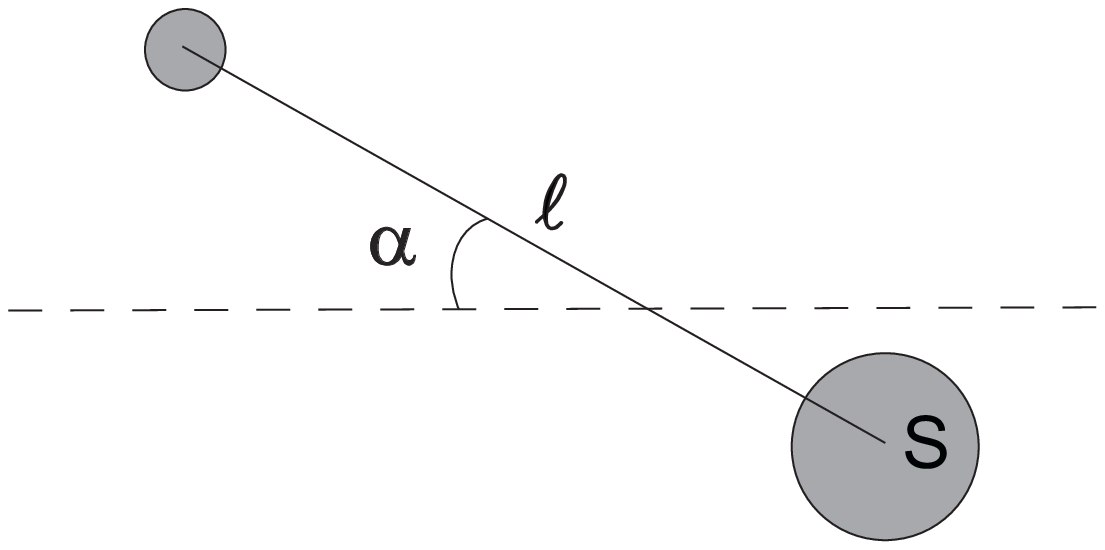}}
 \begin{description}
 \item{\small Fig.~1. Illustration of Joy’s law and parameters of Eq. (1).
The dashed line indicates the local solar parallel. The Northern
Hemisphere, the pole is on the top.
    }
 \end{description}
\end{figure}

Since the tilt angle $\alpha$ is finite, the magnetic field of
active regions contains a poloidal component. During the decay of
active regions and subsequent turbulent diffusion over the solar
surface, this component will contribute to the Sun’s global poloidal
field. We believe that the total contribution of active regions in
an individual solar cycle to the poloidal field is proportional to

\begin{equation}
    B = \sum\limits_{i}^{} S_i \ell_i \sin\alpha_i ,
    \label{1}
\end{equation}
where the summation is over all active regions and the summable
quantities are taken during the maximum development of the sunspot
group. The quantities on the right-hand side of the formula are
explained in Fig.~1: $S_i$ is the area of the largest sunspot in the
group, $\ell_i$ is the separation between the centers of opposite
polarities, and $\alpha_i$ is the tilt angle.

Let us explain Eq.~(1). The contribution to the poloidal field is
assumed to be proportional to the magnetic flux from an active
region. Since the magnetic field strength in developed sunspots
varies within a moderately wide range, approximately from $2.5$ to
$3.5$~kG (Obridko 1985), the magnetic flux may be considered to be
proportional to the area of the largest spot. The characteristic
separation between group sunspots is small compared to the solar
radius. Therefore, after the group decay, turbulent diffusion will
lead to the "cancellation"\ of opposite polarities. Only a small
part of the magnetic flux from an active region will contribute to
the global poloidal field. Babcock (1961) estimated this
contribution to be about 1\%. One may expect this contribution to
increase with $\ell$. Since $\ell$ is small (compared to $R_\odot$),
the dependence on this quantity in Eq.~(1) is assumed to be linear.
Finally, the poloidal field component in an active region is
proportional to $\sin\alpha$.

Equation~(1) includes the minimal set of parameters to estimate the
Babcock–Leighton mechanism. Nevertheless, we managed to find only
one long series of (digitalized) sunspot data suitable for such an
estimation. The Catalog of Solar Activity (CSA) of the Pulkovo
Astronomical Observatory (http://www.gao.spb.ru
/database/csa/groups\_r.html; Nagovitsyn et al. 2008) provides this
possibility. The data of this catalog allow $B$~(1) to be calculated
for three solar cycles.

 \bll
 \centerline{RESULTS}
 \bl

Figure~2 shows the values of $B$ of Eq.~(1) for solar cycles 19–21
and the $A$-index of the large-scale field (Makarov and Tlatov 2000)
for the solar minima following these cycles. Also shown here are the
positions of these three cycles in coordinates of $A$ and the
so-called solar cycle amplitude $W$ ($W$ is the maximum value of the
yearly mean Wolf numbers).

It has been repeatedly pointed out, and we confirm this conclusion,
that the quantities $A$ and $W$ do not correlate with each other
(Makarov and Tlatov 2000; Makarov et al. 2001; Jiang et al. 2007).
In terms of the dynamo theory, this implies that there is no
functional relationship between the toroidal field at the cycle
maximum and the poloidal field at the succeeding minimum. This
relationship is believed to be realized by the $\alpha$-effect and
its ambiguity is due to random fluctuations inherent in this effect.

\begin{figure}[htb]
 \centerline{
 \includegraphics[width=12cm]{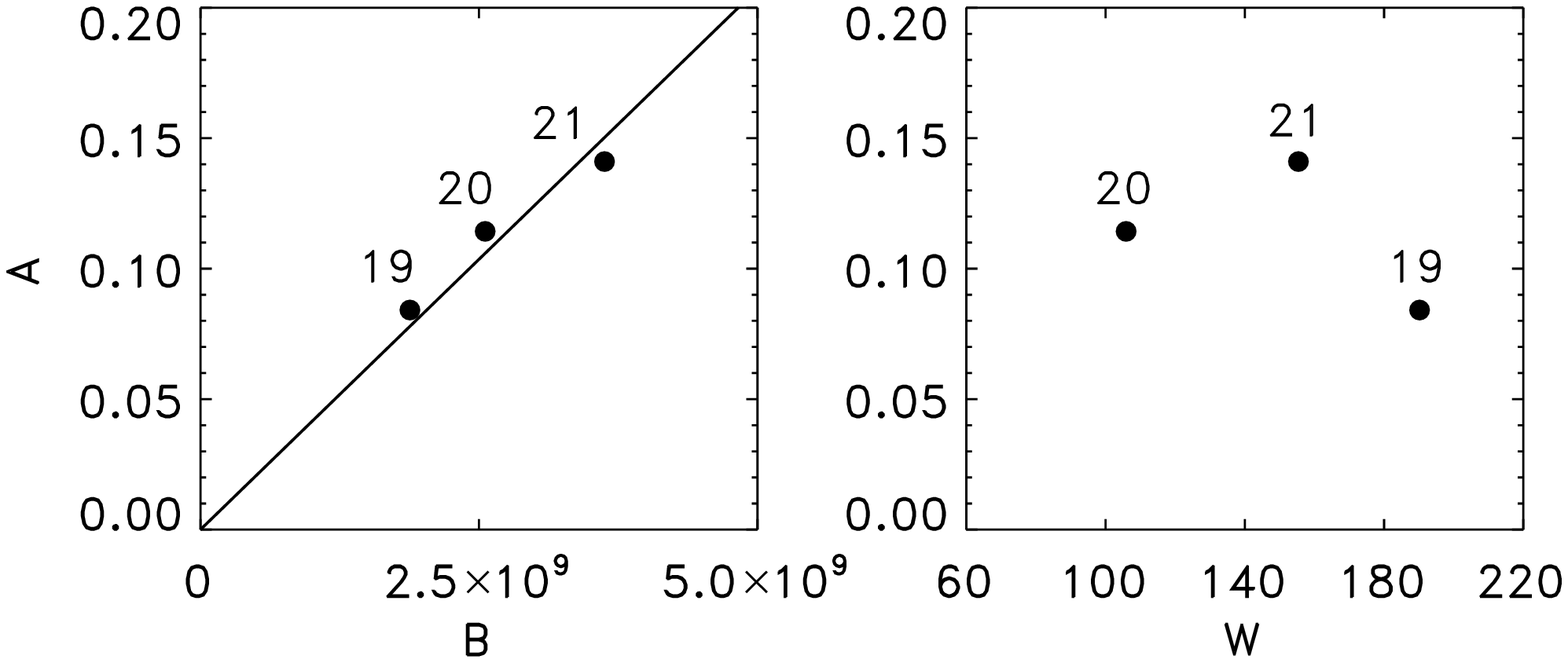}}
 \begin{description}
 \item{\small Fig.~2. {\sl Left}: positions of individual solar cycles in coordinates of
      $B$ calculated from Eq.~(1) and $A$-index at the succeeding activity
      minimum. {\sl Right}: the same for the amplitude $W$ of Wolf numbers and the
      $A$-index.
     }
 \end{description}
\end{figure}

However, this does not rule out the possibility of estimating the
contribution of the $\alpha$-effect with all the existing
fluctuations to the poloidal field generation. Equation~(1) gives
such an estimate for a special variety of the $\alpha$-effect that
is commonly called the Babcock–Leighton mechanism. The values of $B$
obtained from this estimate closely correlate with the $A$-index
(Fig.~2). The line in the left part of the figure represents the
relation
\begin{equation}
    A = 4.1\times 10^{-11} B ,
    \label{2}
\end{equation}
where in calculating $B$ from Eq.~(1), we expressed $S_i$ in
millionths of the solar hemisphere and $\ell_i$ in kilometers. Of
course, the calculations for only three solar cycles do not allow us
to judge the reliability of the relation found with full confidence.
Nevertheless, the results of Fig.~2 indicate that the
Babcock–Leighton mechanism actually operates on the Sun.

\bl \centerline{ACKNOWLEDGMENTS}

This work was supported by the Russian Foundation for Basic Research
(project nos 10-02-00148 and 10-02-00391).
\bl
\setlength{\baselineskip}{0.5 truecm}

\centerline{REFERENCES}
\begin{description}

\item Babcock,\,H.W.: 1961,
    \apj , {\bf 133}, 572.
\item Howard,\,R., Harvey,\,J.: 1970,
    \sp , {\bf 12}, 23.
\item Jiang,\,J., Chatterjee,\,P., Choudhuri,\,A.R.: 2007,
    \mnras , {\bf 381}, 1527.
\item Kitchatinov,\,L.L., Olemskoy,\,S.V.: 2011a,
    Astron. Lett., {\bf 37}, 286.
\item Kitchatinov,\,L.L., Olemskoy,\,S.V.: 2011b,
    \an , {\bf 332}, 496.
\item Labonte,\,B.J., Howard,\,R.: 1982,
    \sp , {\bf 75}, 161.
\item Leighton,\,R.B.: 1969,
    \apj , {\bf 156}, 1.
\item Makarov,\,V.I., Tlatov,\,A.G.: 2000,
    Astron. Rep., {\bf 44}, 759.
\item Makarov,\,V.I., Tlatov,\,A.G., Kallebaut,\,D.K., et al.: 2001,
    \sp , {\bf 198}, 409.
\item Moss,\,D., Sokoloff,\,D., Usoskin,\,I., Tutubalin,\,V.: 2008,
    \sp , {\bf 250}, 221.
\item Nagovitsyn,\,Yu.A., Miletskii,\,E.V., Ivanov,\,V.G.,
    Guseva,\,S.A.: 2008,
    Kosmich. Issled., {\bf 46}, 291.
\item Newton,\,H.W., Nunn,\,M.L.: 1951,
    \mnras , {\bf 111}, 413.
\item Obridko,\,V.N.: 1985,
    {\it Sunspots and Activity Complexes}, Nauka, Moscow (in Russian).
\item Parker,\,E.N.: 1955,
    \apj , {\bf 122}, 293.
\item Parker,\,E.N.: 1979,
    {\it Cosmical Magnetic Fields}, Clarendon Press, Oxford.
\item Schatten,\,K.H., Scherrer,\,P.H., Svalgaard,\,L., et al.:
    1978,
    Geophys. Res. Lett., {\bf 5}, 411.
\item Schmitt,\,D.: 1987, \aa , {\bf 174}, 281.
\item Schou,\,J., Antia,\,H.M., Basu,\,S., et al.: 1998,
    \apj , {\bf 505}, 390.
\item Steenbeck,\,M., Krause,\,F., R\"adler,\,K.-H.: 1966,
    Z.~Naturforsch., {\bf 21a}, 369.
\item Vainshtein,\,S.I., Zel’dovich,\,Ya.B., Ruzmaikin,\,A.A.: 1980,
    {\it Turbulent Dynamo in Astrophysics}, Nauka, Moscow (in Russian).
\item Wang,\,Y.-M., Sheeley,\,N.R. Jr.: 1989,
    \sp , {\bf 124}, 81.
\item Wilson,\,P.R., Burtonclay,\,D., Li,\,Y.: 1997,
    \apj , {\bf 489}, 395.
\end{description}

\bl

\hspace{9truecm}{\sl Translated by G.~Rudnitskii}

\end{document}